\documentclass[10pt,twocolumn]{article}

\usepackage[utf8]{inputenc}
\usepackage[english]{babel}
\usepackage{amsmath}
\usepackage{amsfonts}
\usepackage{amssymb}
\usepackage{mathpazo} 
\usepackage[scaled]{helvet}
\usepackage[T1]{fontenc}
\usepackage{url}
\usepackage{verbatim}
\usepackage{float}
\usepackage{caption}
\usepackage{subcaption}
\usepackage{lineno}
\usepackage{graphicx}
\usepackage{xcolor}
\usepackage{booktabs}
\usepackage{algorithm}
\usepackage[noend]{algpseudocode}
\usepackage{changepage}
\usepackage{svg}
\usepackage{soul}
\usepackage[left=48pt,%
                right=42pt,%
                top=42pt,%
                bottom=60pt,%
                headheight=15pt,%
                headsep=10pt,%
                letterpaper,twoside]{geometry}%
                
\usepackage[labelfont={bf,sf},%
                labelsep=period,%
                figurename=Fig.,%
                singlelinecheck=off,%
                justification=RaggedRight]{caption}
                
\usepackage[numbers,sort&compress]{natbib}
\title{Single-Shot, Spatio-Temporal Metrology of Relativistic Plasma Optics}

\usepackage{authblk}
\author[]{Ankit Dulat}
\author{Amit D. Lad}
\author{C. Aparajit}
\author{Anandam Chaudhary}
\author{Yash M. Ved}
\author[2]{Laszlo Veisz}
\author[1,*]{G. Ravindra Kumar}

\affil[1]{Tata Institute of Fundamental Research, 1 Homi Bhabha Road, Colaba, Mumbai 400 005, India}
\affil[2]{Department of Physics, Umeå University, SE-90187 Umeå, Sweden}
\affil[*]{E-mail: grk@tifr.res.in}

\date{}

\begin{document}

\twocolumn[
  \begin{@twocolumnfalse} 
    
    \maketitle
    
    \begin{abstract}
    Ultrahigh peak power femtosecond laser pulses create extreme states of matter that are currently being probed with great interest. Plasma optics have been proposed for shaping and amplifying high-power pulses, but they are subject to huge modulations and fluctuations due to the very nature of excitation at high intensities. Multidimensional characterization (spatial and temporal) of relativistic plasma dynamics is therefore crucial to understand the spatio-temporal structure of intense femtosecond pulses shaped by plasma optics. This is, however, extremely difficult to achieve, particularly at the low repetition rates typical at 100s terawatt to petawatt powers. Here, we present a single-shot, three-dimensional (3D) spatio-temporal and spatio-spectral measurement of such pulses based on spectral interferometry. We reconstruct the 3D temporal structure of the laser pulse $\it simultaneously$ resolving the complex plasma dynamics. We demonstrate our method by measuring the sub-picosecond evolution of relativistic solid-density plasmas. Our measurements reveal that different spatial regions of the plasma surface move differently yet exhibit a collective behavior globally. This all-optical measurement technique captures 3D spatio-temporal effects within pulses with ultrahigh peak powers, all in a single shot, enabling further progress in ultrahigh-intensity laser and plasma technologies.
    
   \vspace{2mm}
    \end{abstract}
    
  \end{@twocolumnfalse}
]

\section{Introduction}
Recent years have witnessed a surge of interest in plasma optics, an emerging field focused on the manipulation of high-intensity light pulses spanning picoseconds to attoseconds \cite{Veisz2021,Dromey2023}. The adoption of plasma as an alternative to conventional optics \cite{Thaury2007,Dromey2007,Ren2007,Marques2019,Leblanc2017,Kong2017} presents a compelling solution to the challenges associated with optical damage -- a key factor currently influencing the size and costs of large-scale laser facilities.

Substantial work has been done both theoretically and experimentally to demonstrate the operational principles of plasma-based optical elements. Examples include plasma mirrors \cite{Thaury2007}, which have been used to improve the contrast of laser beams and generate attosecond light pulses \cite{Dromey2007, Quere2006,Kormin2018}, laser amplifiers \cite{Ren2007, Marques2019,Vieux2023}, laser compressors \cite{Malkin1999,Hur2023}, plasma holograms \cite{Leblanc2017, Kong2017}, frequency converters \cite{Peng2021}, plasma lenses \cite{Svensson2021}, and plasma wave plates \cite{Turnbull2016}. Plasma gratings \cite{Monchoce2014} are routinely used at the National Ignition Facility to tune the implosion symmetry of ICF targets by facilitating power transfer between intense lasers \cite{Glenzer2010, Moody2012}. However, there has been a marked scarcity of experimental characterizations of these plasma-optical elements. The key question is: Does the spatio-temporal structure of these femtosecond (fs) pulses remain unaffected after their interaction with the plasma?

In contrast to conventional optical elements with flat or well-defined shapes, plasma is inherently dynamic, highly complex, and challenging to control. Furthermore, it can be modified by the interaction with the intense fs laser pulses and exhibit strong nonlinear phenomena including instabilities. These can introduce significant spatio-temporal coupling (STC) effects \cite{Jolly2020} in the resulting laser beam or its generated high-order harmonics. These coupling effects are known to have pronounced impacts not only on pulse duration \cite{Jolly2020, Dorrer2019} and focal spot size but also on more intricate aspects such as pulse contrast \cite{Nomura2007}. Hence, it is crucial to comprehensively characterize the complex plasma dynamics and the STC effects they induce in laser pulses. Such characterization is not only essential for alleviating plasma-induced STC effects in laser pulses, but also offers the ability for in-situ control over complex plasma dynamics.

However, this measurement is very challenging, mainly due to two key factors: (1) It requires the simultaneous measurement of ultrafast plasma dynamics during the pulse interaction and its subsequent impact on the spatio-temporal structure of the pulse. (2) It must be a single-shot measurement. The latter requirement arises due to inherent shot-to-shot fluctuations and low repetition rates typical of high-power laser systems.

Significant efforts have previously been devoted to the complete characterization of the three-dimensional (3D) optical field of laser pulses. These include multi-shot techniques that require scanning spatial positions \cite{PamelaBowlanPabloGabolde2007} or temporal delays \cite{Pariente2016, Borot2020}. More recently, single-shot methods \cite{Tang2022,Xu2022,Goldberger2022} based on iterative phase retrieval have been developed, but require complex setups and heavy post-processing algorithms. The other methods \cite{Gabolde2006,Cousin2012} suffer mainly from limited spectral resolution, measuring only a few wavelength channels and relying on interpolation and extrapolation to reconstruct the complete spectrum and spectral phase.

Despite the impressive progress, these methods have rarely \cite{Leblanc2016,Chopineau2021} been applied to characterize plasma-induced STC effects in laser pulses or their high harmonics due to the challenge of implementing such techniques in the extremely harsh conditions of laser-plasma experiments \cite{Leshchenko2019}. Instead, they have predominantly addressed issues related to the misalignment of stretchers and compressors in high-power laser systems or the correction of chromatic aberrations introduced by external optical elements.

In this article, we propose and demonstrate an all-optical method for single-shot, 3D spatio-temporal characterization of fs pulses interacting with relativistic hot-dense plasma. We further demonstrate how the measured spatio-temporal and spatio-spectral effects in the laser pulses can be utilized to deduce the "complete" evolution of plasma on fs timescales with a few micrometer spatial resolution. Our method is simple, experimentally robust, and is readily implemented or upgraded in a laser-plasma laboratory setting.

\begin{figure*}[!ht]
\centering
\includegraphics[width=\linewidth]{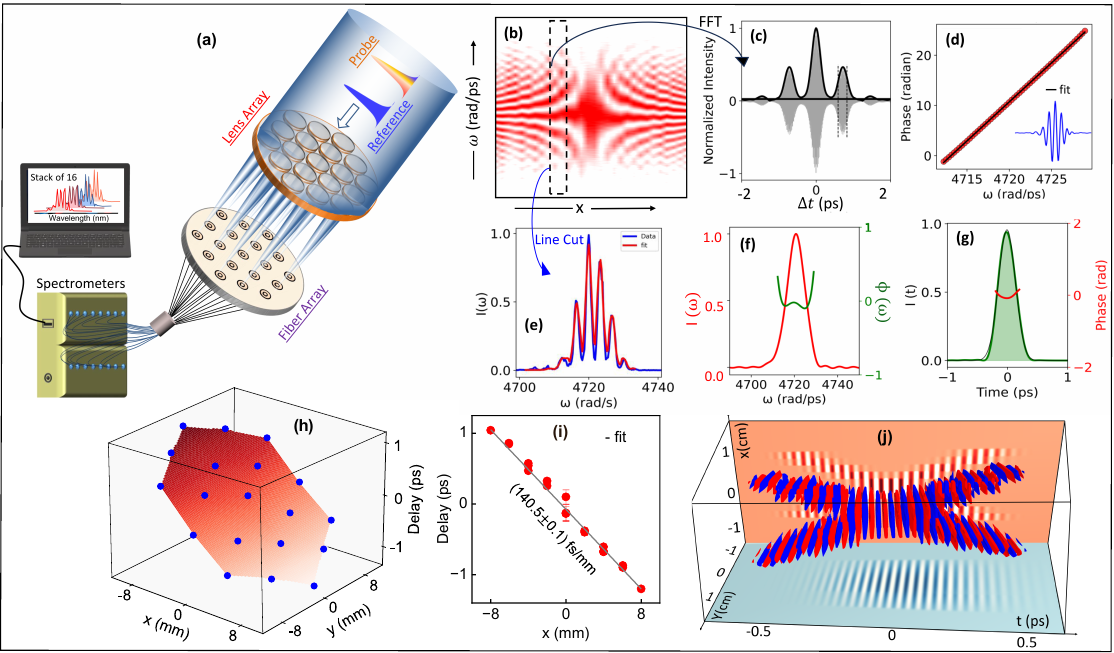}
\caption{\textbf{Working principle and reconstruction procedure for single-shot STC measurement. a} A well-defined optical field is divided into two time-delayed copies, one of which acts as a reference (R), while the other serves as a probe (P), whose spatio-temporal structures need to be determined. Subsequently, the spatial profiles of these copies are sampled using a lens array, which focuses distinct spatial segments of light onto a fiber array, which couples light to a stack of spectrometers. This enables us to capture spectral fringes across multiple spatial locations simultaneously, all in a single shot. \textbf{b-g}, Fourier analysis for the reconstruction of $\varphi$(x,y,$\omega$) and temporal profile: Contour plot of raw spectral fringe data sampled by the lens array, where each vertical slice corresponds to distinct spatial locations (\textbf{b}). \textbf{c} Fourier transform of a vertical slice of the raw data. Inset in (\textbf{d}) shows the inverse Fourier transform (IFFT) of the filtered side-peak in (\textbf{c}). The imaginary part of the logarithm of IFFT (\textbf{d}) represents the spectral phase ($\varphi (\omega)$). \textbf{e} Reconstructed spectral fringes from the measured $\varphi (\omega)$ with the frequency-independent phase $\varphi_0$ (red); $\varphi_0$ is obtained from fitting the raw data (blue). Reconstructed spectral and temporal profiles are depicted in \textbf{f} and \textbf{g} respectively. \textbf{h} The measured pulse front delay across the beam and corresponding interpolated surface. \textbf{i} The x cross-section of data points in \textbf{h} with linear fit. \textbf{j} An isosurface plot illustrates the complete spatio-temporal electric field distribution of the measured light pulse. To clearly show the peaks and valleys of the optical fields in \textbf{j}, the carrier frequency was reduced 40 times and plotted with an amplitude threshold of 0.4. Red and blue colors correspond to the positive and negative values of the electric field.}\label{fig: Fig1}
\end{figure*}

\vspace{-4mm}
\section{Results}

\subsection{Technique}

The measurement method employed in the experiment is an extension of a powerful technique known as Spectral Interferometry (SI) \cite{Tokunaga1992}. Unlike iterative phase-retrieval based methods \cite{Tang2022,Xu2022,Goldberger2022}, this technique involves a much simpler setup, is more direct and considerably faster. Additionally, since it is a spectrum-based measurement, it allows the use of high-resolution spectrometers, proving especially useful for characterizing plasma dynamics (as discussed later). Despite considerable development \cite{Oksenhendler2010,OKSENHENDLER2017,Alonso2012} over the years, this method hasn't been adapted for single-shot, 3D spatio-temporal characterization of laser pulses due to the challenge of capturing 3D information (two spatial dimensions plus one spectrum) on a 2D/1D detector.

We addressed this issue by spatially sampling the laser beam in the transverse direction and simultaneously examining each sample with a different detector, using a common 
reference pulse for all the samples. This reference allows us to compare the properties of laser pulses across the entire beam. The measurement device is sketched in Fig. \ref{fig: Fig1}a, and simply consists of a lenslet array and a fiber array coupled with a stack of spectrometers. We measure the spatio-temporal profile of the unknown or probe pulse (P) via its spectral interference with a time-delayed, well-defined reference pulse (R). The spatial profile of both reference and probe pulses is sampled by the lens array, which focuses distinct spatial segments of light onto a fiber array, which then couples light to a stack of sixteen individual spectrometers. This enabled the simultaneous capture of spectral interference across multiple spatial locations, all in a single shot. The setup is very simple, highly robust, and insensitive to laser fluctuations and shot-to-shot vibrations owing to the collinear paths of both the reference and probe. The setup is exceptionally well-suited for capturing plasma-induced STC effects and characterizing light pulses from high-power laser systems. The setup can be further simplified by utilizing an imaging spectrometer and positioning the fibers from the array at the entrance slit of the spectrometer.

\subsection{STC characterization of fs pulses}
First, to demonstrate the working principle of the technique we intentionally introduced complex spatio-temporal coupling to the second harmonic light pulses of a Ti-sapphire, 800 nm, femtosecond laser pulse (400 nm, 260 fs pulse width). A Fresnel biprism was placed in the collimated pulse path (see Figure S1 in the Supplementary Information for more details); it introduced angular dispersion in the two spatial halves of the beam along the x-direction, which resulted in a relative pulse front tilt of 140 fs/mm between the two halves of the beam. One half served as a reference and the other as the probe pulse. Fig. \ref{fig: Fig1}(b-j) depicts the reconstruction procedure of the 3D optical field of reference and probe pulse in the intersection plane of two spatial halves.

The raw data sampled by the lenses in the lenslet array is presented in Fig. \ref{fig: Fig1}b as a contour plot. Each vertical slice in the plot corresponds to spectral fringes measured at different spatial positions along the x-axis. The fringe spacing and number of fringes vary along the x-direction, indicating spatially varying delay between pulse halves, with zero delay at the center and increasing delay towards the edges due to the inverse proportionality of fringe spacing to delay. The spectral phase $\varphi (\omega,\textbf{r})$ is retrieved for each slice in Fig. \ref{fig: Fig1}b via a fast Fourier transform (FFT) based algorithm \cite{Takeda1982} (see Methods). To explain the crucial effects of the spectral phase on an ultrashort laser beam, we use a Taylor expansion with respect to frequency around the central frequency $\omega_0$ of the pulse:
\begin{equation} \label{eq:1}
    \varphi(\omega,\textbf{r}) = \varphi(\omega_0,\textbf{r}) + \frac{\partial\varphi}{\partial\omega}\delta\omega + \frac{1}{2}\frac{\partial^2\varphi}{\partial\omega^2}\delta\omega^2+\Phi(\omega,\textbf{r})
\end{equation}
where $\delta\omega = \omega$ – $\omega_0$ and $\Phi(\omega,\textbf{r})$ includes all terms of the Taylor expansion of order higher than 2 in $\delta\omega$. The first term, $\varphi(\omega_0,\textbf{r})$, describes the carrier-envelope phase profile, and the second term, $\tau(\textbf{r})$= $\partial\varphi/\partial\omega$ is the local group delay of the pulse and determines the arrival time of the light pulse at position r. The three-dimensional surface $\Sigma$=(\textbf{r},z = c$\tau$(\textbf{r})) represent the pulse front of the beam. The third term represents the temporal chirp and can affect the pulsewidth of the pulse. The last term $\Phi(\omega,\textbf{r})$ accounts for the higher-order distortion of the temporal shape of the pulse and its possible variations across the beam. The measured group delay across the beam and corresponding interpolated surface are shown in Fig. \ref{fig: Fig1}h, and the corresponding x cross sections of the data points are shown in Fig. \ref{fig: Fig1}i, which shows a relative tilt of 140.5 $\pm$ 0.1 fs/mm between the two halves of the beam, which is in good agreement with the expected theoretical value of 140 fs/mm. The isosurface plot of the reconstructed spatio-temporal profile of the total field i.e. reference plus probe is depicted in Fig. \ref{fig: Fig1}j. 
The pulse duration at different spatial locations closely resembles the undispersed pulse duration of $\sim$260 fs. However, the effective global pulse width has become much larger ($\sim$1 ps) due to the large pulse front tilt in the two halves of the pulse. 

\subsection{Application to study the femtosecond evolution of hot, dense plasma}

Here, we present an important application of this all-optical measurement scheme to investigate the in-situ interaction dynamics of a relativistic laser pulse with solid-density plasma. Such interactions facilitate high-energy density (HED) science and offer high brightness, ultrafast particles, tabletop-scale XUV light sources \cite{Rajeev2003}, and plasma optics \cite{MURNANE531,Riconda2023}. Despite the crucial need for understanding the 3D spatio-temporal dynamics inherent to these interactions, direct measurement of their evolution on fs timescales remains a formidable challenge.  

\begin{figure}[ht]
\centering
\includegraphics[width=1.1\linewidth]{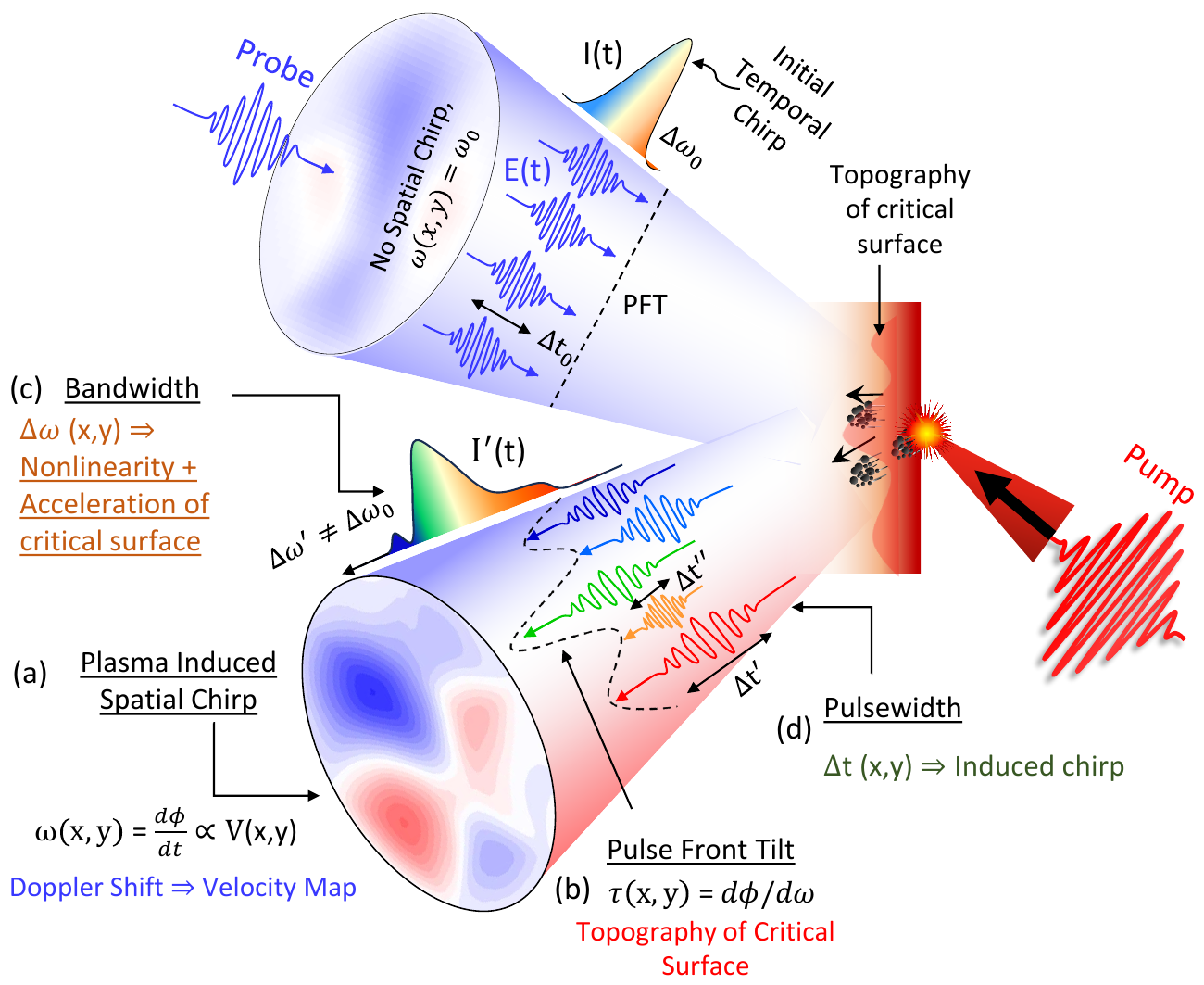}
\caption{\textbf{Application of spatio-temporal and spatio-spectral effects in capturing the dynamics of hot, dense plasma.} STC characterization, in a plane where the target surface is imaged, combined with the pump-probe technique captures in-situ laser-induced plasma dynamics at the rear of a thin target. A probe pulse, initially with a small temporal chirp, interacts with laser-produced plasma. The pulse front tilt (PFT) of the reflected probe maps critical surface topography. Spectral chirp in the probe pulse maps the 2D velocity of the reflecting surface, while spatial modulation in the probe pulse bandwidth reveals critical surface acceleration.}\label{fig: cartoon}
\end{figure}

The schematic in Fig. \ref{fig: cartoon} discusses how different spatio-temporal and spatio-spectral effects induced in the fs probe pulse during the interaction can be utilized to infer these complex dynamics. To gain a deeper understanding, let us carefully examine each of the effects. The probe pulse, with a flat pulse front and a small temporal chirp, interacts with the critical surface in the dense plasma. Its temporal and spectral phases are modulated by the plasma differently across its spatial profile. These modulations can introduce a significant spatial chirp, affect the curvature of the pulse front, and introduce a strong modulation in the spectral and temporal profile of the pulse, as depicted in Fig. \ref{fig: cartoon}. The temporal phase experienced by the probe pulse is,
\begin{equation} \label{eq:2}
\begin{split}
\phi(t,\textbf{r}) & = 2\int^{z=z_{c}(t)}_{z_R} \frac{\omega}{c}\sqrt{\epsilon(\textbf{r},s)} ds \\
 & \approx \frac{2\omega_0}{c}\left (\int^{z=z_{c}(t)}_{z_R}ds - \frac{1}{2n_{c}}\int^{z=z_{c}(t)}_{z_R}n_{e}(\textbf{r},s,t)ds\right)\\
  & \hspace{15 mm}+  \psi[f(n_e/n_c)]
\end{split}
\end{equation}

where $z_R$ is a reference point far from the target, $z_c(t)$ is the position of the reflecting surface at time t, \textbf{r} is the coordinate along the transverse spatial profile and s is along the beam path. $\epsilon$ is the permittivity that can be approximated as $\epsilon$=1-(n$_{e}$/n$_{c}$), with  n$_{e}$ and n$_{c}$ being the local electron density and critical density, respectively. $\psi$ is the function consisting of the integral of higher-order terms in the Taylor series expansion of $\sqrt{\epsilon}$.

The imaginary part in $\epsilon$ has been neglected since it mainly causes absorption of the probe pulse, contributing weakly to the phase shift. The induced temporal phase modulation given in Eq. \ref{eq:2} can strongly affect the spectral profile of the pulse that interacts with the plasma. The first term represents a linear phase shift induced by the movement of the reflecting surface due to hydrodynamic expansion or motion of the ionization front. This leads to a shift in the central frequency of the spectrum via Doppler shift \cite{Landen1992,Adak2015,Mondal2010}
\begin{equation} \label{eq:3}
    \omega(\textbf{r},t_0)-\omega_0= \frac{d\phi_1(\textbf{r},t)}{dt} = 2\omega_{0} V(\textbf{r})cos(\theta)/c
\end{equation}
where $\phi_1(\textbf{r},t)$ is the first term in Eq. \ref{eq:2}, $V(\textbf{r})$ is the velocity of the reflecting surface at spatial position \textbf{r} and $\theta$ is the incidence angle of the probe.

The position-dependent shift in the central frequency of the pulse induces a spatial chirp in the reflected pulse, which can be used to reconstruct the 2D velocity map of the reflecting surface. The high spectral resolution (0.03 nm) offered by our technique enables the measurement of velocities from as low as $10^4$ m/s, up to relativistic motion, such as that in a relativistic oscillating mirror (ROM) in high-harmonic generation (HHG).\\

\begin{figure*}[!ht]
\centering
\includegraphics[width=1\linewidth]{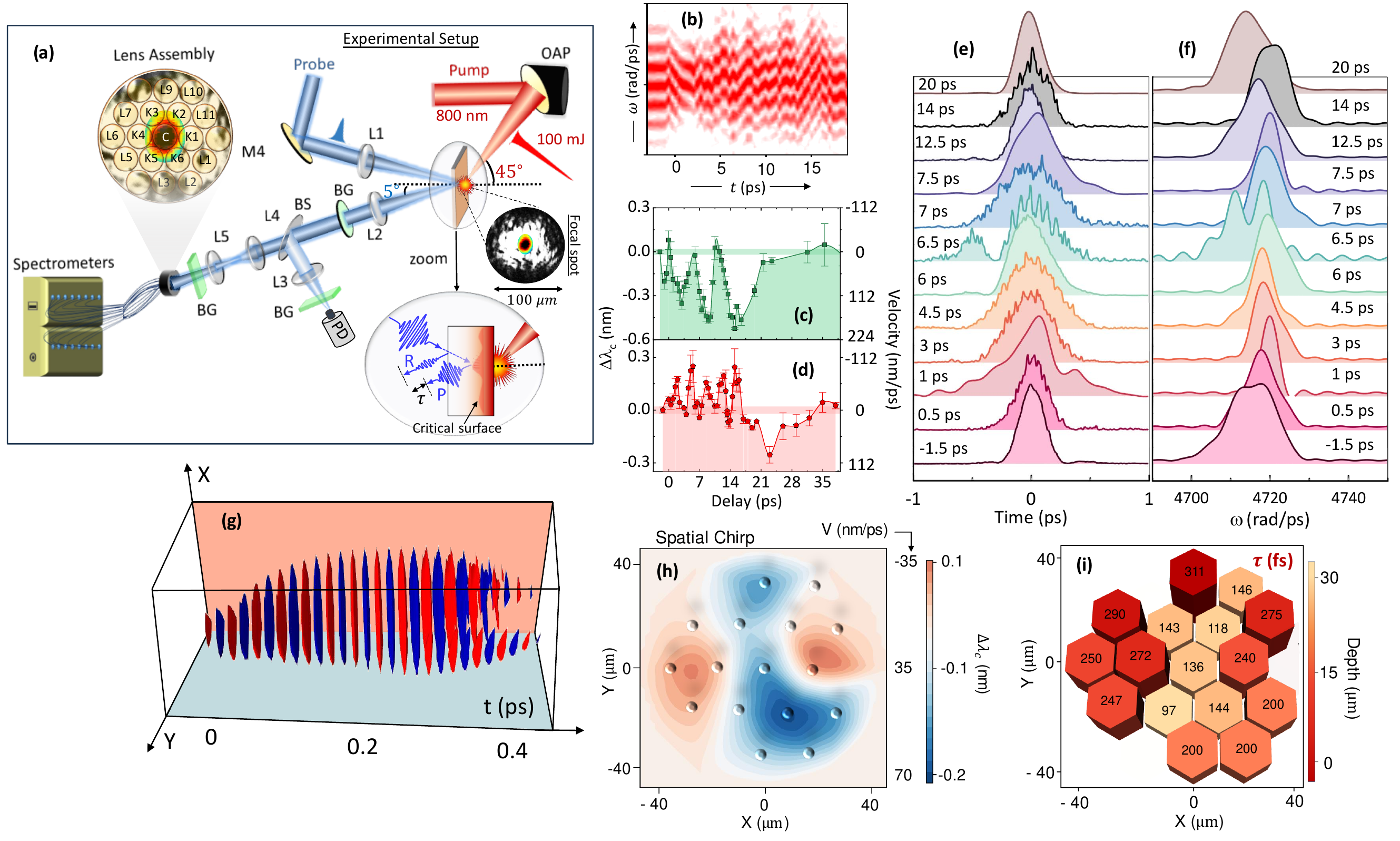}
\caption{\textbf{Probing femtosecond evolution of hot, dense plasma by characterizing plasma-induced STC effects. a} Schematic of pump-probe experiment. The front side of the copper-coated FS target is irradiated by a pump pulse (800 nm, 100 mJ, 25 fs), while a second-harmonic (SH) probe pulse captures the plasma generated on the copper-glass interface from the rear side. The weak reflection from the glass-vacuum interface acts as a reference, and the reflection from the plasma acts as the probe. The probe focal spot on the target is imaged on the lens array of the STC setup. The inset depicts the measured focus of the pump (in color) and probe pulse (in black and white) on the target. \textbf{b} Contour plot of the spectral fringes captured by the K4 lens as the probe is delayed with respect to the pump. (\textbf{c,d}) Compares the evolution of the induced Doppler shift and the corresponding propagation velocity of the reflecting surface at two different spatial locations, K3 and K4, respectively. (\textbf{e,f}) Depiction of the changes in the temporal intensity and spectral profile of the probe pulse as the plasma evolves over different time delays at location K4. \textbf{g} An isosurface plot of the spatio-temporal structure of the electric field of the probe pulse at a delay of 0.5 ps, where red and blue colors correspond to positive and negative values of the electric field. \textbf{h} Contour plot of the fitted spatial chirp obtained from the data measured at the points indicated as dots at a delay of 0.5 ps and the corresponding propagation velocity of the reflecting surface. \textbf{i} Hexagonal bar plot of the group delay measured across the spatial profile of the pulse at a delay of 0.5 ps and the corresponding depth of the reflecting plasma surface; each hexagonal bar represents the data sampled by each lens in the lens array. To clearly show the peaks and valleys of the optical fields in \textbf{g}, the carrier frequency was reduced 25 times and plotted with an amplitude threshold of 0.5.}\label{fig: Fig3}
\end{figure*}

Besides Doppler shifts, the first term in Eq. \ref{eq:2} can also change the bandwidth ($\Delta\omega$) of the reflected probe, that is, it can increase or decrease. This occurs when the probe pulse experiences a quadratic phase modulation over time. One way this modulation can be induced is when the reflecting surface is not moving at a constant velocity, but has an acceleration. As a result, different frequency components of the initially chirped probe pulse experience varying Doppler shifts due to the acceleration of the reflecting surface. (see Section S3 and Fig. S6 in the Supplementary Information (SI) for more details.) Observing whether the bandwidth of the original slightly chirped pulse shrinks or increases allows the determination of a positive or negative acceleration. Additionally, the exact magnitude of the acceleration can be calculated given knowledge of the initial chirp in the probe pulse \cite{Sauerbrey1996}. Thus, our single-shot measurement technique is also capable of capturing the 2D acceleration map of the reflecting plasma surface.

The second and higher order terms in Eq. \ref{eq:2} represent the nonlinear phase modulation induced by rapid modulation of the electron density profile, which is often neglected, since contributing weakly during the rapid hydrodynamic expansion of the plasma (see Fig. S6 in Supplementary Information). However, these terms can make a significant contribution when the density profile changes rapidly, such as in the early few femtoseconds of rapid ionization by the pump pulse \cite{Dragila1987}. Such a nonlinear temporal phase can induce very complex modulations in the spectral profile of the pulse, ranging from broad humps to intricate structures in the initial Gaussian profile. We show that contributions from these higher-order terms are significant even after several picoseconds in the relativistic plasma dynamics inside solids and hence need to be considered. Our method allows the measurement of these complex spectral modulations in the reflected probe pulse very conveniently, which can be utilized to gain insight into the nonlinear evolution of the electron density profile \cite{Dulat2022}.

Having discussed spatio-spectral effects, let us delve into how spatio-temporal effects arise in the probe pulse and the insights they provide for inferring plasma dynamics. Referring back to Eq. \ref{eq:1}, the first-order term in the spectral phase is the group delay. Therefore, when the probe pulse with a flat pulse front (i.e. $\tau(\textbf{r})$ is the same across its transverse profile) interacts with the critical surface of the plasma, which has some topography as sketched in Fig. 2, then the different segments of the pulse will experience a different amount of group delay and hence will induce curvature in the pulse front. This pulse front curvature (PFC) can be gauged very conveniently in our measurement scheme by assessing the local group delay across the transverse spatial profile. Using this, we can construct the topography of the critical surface. Having a common reference pulse, which serves as a baseline for comparing the group delay of all segments of the reflected probe pulse, is crucial and is indeed an integral part of our setup.

\subsection{Capturing relativistic plasma dynamics}
To demonstrate the concept and adaptability of our method, we combined this with a pump-probe technique. This offers a whole new possibility of measuring plasma dynamics, from a few fs time delays to several picoseconds (ps) after the interaction.

We measured the complete 3D evolution of the relativistic plasma driven by energetic electrons deep inside a solid target. This investigation is particularly significant due to its inherent complexity, limited understanding, and the absence of direct measurements.

The experimental setup is depicted schematically in Fig. \ref{fig: Fig3}a. P-polarised laser pulses (25 fs, 800 nm) were focused to a 10 $\mu$m spot at an incidence angle of 4$0^{\circ}$, creating a peak intensity of $\sim10^{18}$ W/cm$^2$ (see Methods). We used a Cu-coated FS glass plate as the target in our experiment, with Cu on the front side. This serves two purposes: One, since FS is transparent, the probe from the rear side can go deep inside the glass and interact with the plasma as close as the Cu-glass interface and capture its subsequent evolution within the glass. Second, there is a weak reflection from the vacuum glass interface that does not interact with the plasma; this can serve as a reference pulse for spatio-temporal measurements.

We use the second harmonic (400 nm) of the pump pulse as a probe since using it has multiple advantages. First, it has a narrow spectrum ($\sim$ 1 nm) with a nearly Gaussian profile (unlike pump pulses with bandwidths as large as 60 nm, which have a complex spectral profile), making it easy to characterize even small Doppler shifts. Secondly, it vastly improves the signal-to-noise ratio because all pump noise can be removed with a narrow spectral filter around 400 nm.
The probe was focused on the rear side of the target with a lens (L1) at near-normal incidence ($\sim5^\circ$) to a spot of $\sim$80 $\mu$m. The focal spot size of the probe was several times larger than the pump focal spot, to capture the lateral expansion of the plasma. The reflected probe and reference pulses were collected by an f/2 plano-convex lens and imaged on the lenslet array of the STC measuring device. 

Fig. \ref{fig: Fig3}b shows the contour plot of the spectral fringe data captured by the lens K4 as the probe is delayed from 0 to 20 ps with respect to the pump pulse. The data for each time delay was averaged over several laser shots. As seen in the plot, the fringes are straight just before the pump pulse hits the target, and suffer a strong modulation at later time delays. Aside from the bending and twisting of the fringes, the spectrum's bandwidth is also modulated as the plasma evolves.
In Fig. 3(c,d), we compare the evolution of Doppler shift ($1^{st}$ order modulation in $\phi$(t)) measured at two different spatial locations, K3 and K4. It is very interesting to note that common quasi-periodic oscillations appear in the central frequency of the spectrum for delays as long as 16 ps, which eventually decay at later time delays. More intriguing is the observation of vastly different dynamics, despite the spatial locations are very close to each other. The peak of the Doppler shift and the period of the oscillation appear to be very different in the two cases. It demonstrates how plasma front evolution can be very inhomogeneous locally while still having features of collective dynamics. The corresponding propagation velocity of the reflecting surface is shown on the right axis.

\begin{figure}[t]
\centering
\includegraphics[width=1.1\linewidth]{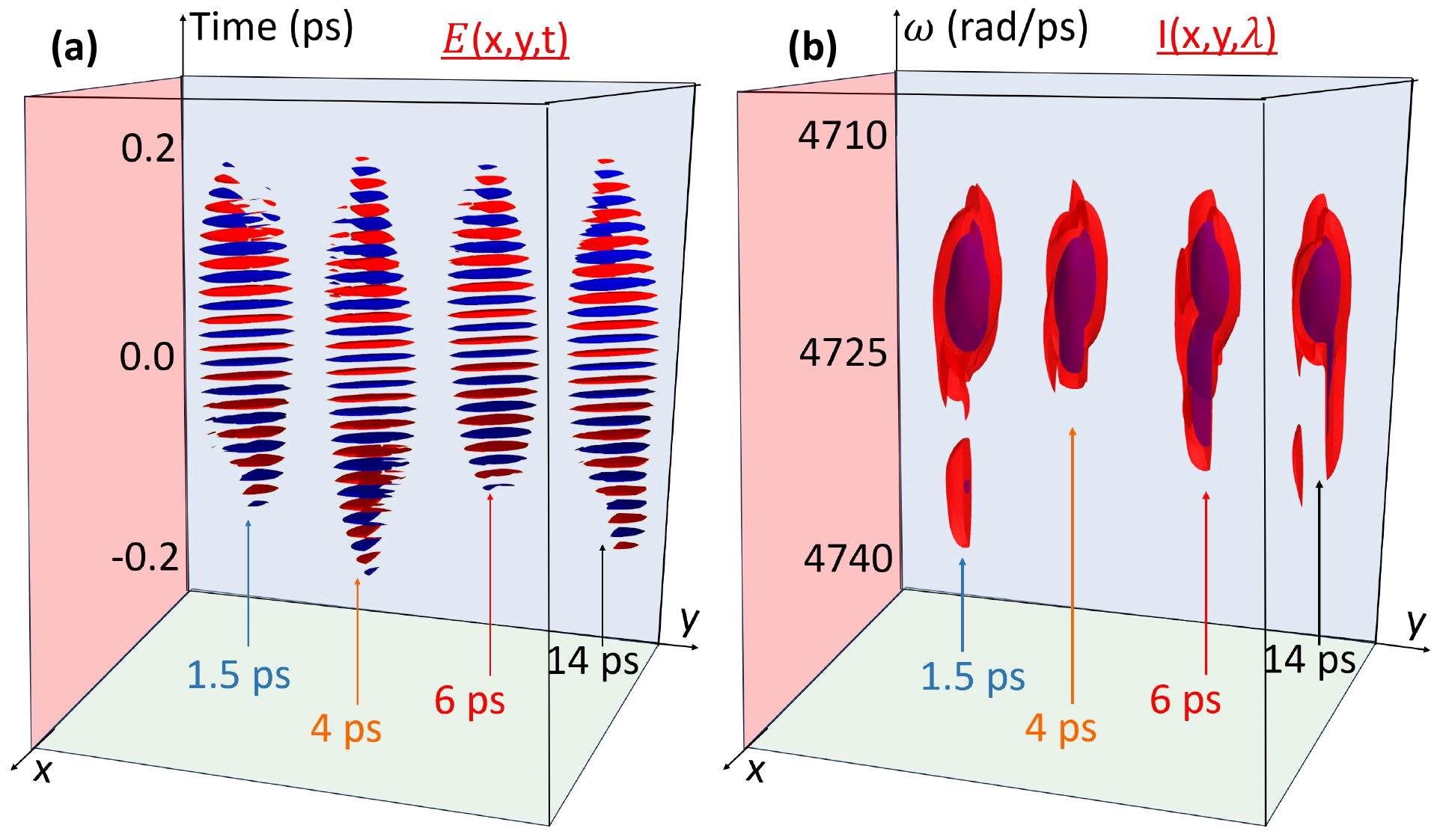}
\caption{\textbf{Plasma induced STC effects in the probe pulse.}  \textbf{a} Isosurface plot illustrating the complete spatio-temporal distribution of the electric field at different time delays. The carrier frequency was numerically reduced by a factor of 25 for enhanced visualization purposes, with an amplitude threshold of 0.5. \textbf{b} Isosurface plot showing sptaio-spectral intensity profile with varying time delay, using amplitude thresholds of 0.3 and 0.6.}\label{fig: Fig5}
\end{figure}

The spectral profile of the probe pulse at various time delays is shown in Figure \ref{fig: Fig3}f, measured at spatial location K4. As can be seen, not only the bandwidth of the probe pulse is changing, but its spectral profile is also getting modulated, and at a delay of about 6.5 ps, it exhibits double peak structures. As previously discussed, electron density profile-induced high-order temporal phase modulation causes this change probably triggered by energetic electrons released during the relativistic laser-plasma interaction. Furthermore, the acceleration of the critical surface causes the bandwidth of the probe pulse to change. The corresponding temporal structure of the probe pulse is plotted in Fig. \ref{fig: Fig3}e. Note that because of higher-order spectral phase modulation induced by the plasma, its temporal structure has been strongly modulated differently at various time delays. Therefore, this characterization becomes very important for the cases where plasma is used as an optical tool for intense femtosecond pulses.

The isosurface plot of the reconstructed electric field of the probe pulse after 0.5 ps of interaction with the plasma is shown in Fig. \ref{fig: Fig3}g. As observed, the effective pulse width has increased from the initial 260 fs, and the plasma has induced curvature in the pulse front, with the central part leading the edges. The curvature is also evidenced in the measured group delay, as shown in Fig. \ref{fig: Fig3}i. It represents that the critical surface is not flat but has a structure, with the central region being ahead of the edges by about 25 µm. The Doppler shift measurement shown in Fig. \ref{fig: Fig3}h provides insight into the instantaneous expansion velocity of the critical surface, revealing that the regions ahead move significantly faster compared to the edges. Using Eq. \ref{eq:3}, we calculated the maximum expansion velocity to be $6\times 10^6$~cm/s and the instantaneous acceleration of $\sim 9\times 10^{17}$~m/s$^2$ (using Eq. 4 in SI) in the central region of the plasma.

We also capture the evolution of the plasma and the spatio-temporal structure of probe pulses over several ps. Fig. \ref{fig: Fig5} compares the spatio-temporal profile and the spatio-spectral profile of the probe pulses over several time delays. As observed in Fig. \ref{fig: Fig5}a, at early time delays of 1.5 ps, the pulse front has a pulse front tilt on the rising edge of the pulse and a complex curvature on the falling edge. At later time delays, not only does the pulse width of the fs pulse change but also these complex pulse front distortions are changing. Similarly, the isosurface plot of the spatio-spectral profile shown in Fig. \ref{fig: Fig5}b shows that at a time delay of 1.5 ps, there is a strong modulation in the spectrum towards the edge compared to the central region. At a time delay of 6 ps, one can see that spectral width has significantly increased over the entire spatial profile.

The capability of our technique to measure these complex spatiotemporal effects paves the way for in-depth exploration and study of the intricate evolution of relativistic plasmas. The detailed physics underlying the dynamics presented here is very interesting and warrants further comprehensive investigation.

\section{Discussion}
In conclusion, we have demonstrated an all-optical method that characterizes relativistic plasma dynamics on femtosecond timescales and their impact on the spatio-temporal structure of femtosecond pulses, all in a single shot. We initially applied this method to reconstruct intentionally induced complex spatio-temporal coupling in a 
femtosecond pulse. By integrating this method with a pump-probe technique, we introduced a way to retrieve the ultrafast dynamics of solid plasmas, leveraging the spatio-temporal and spatio-spectral coupling effects they induce during pulse interaction. This approach enabled us, for the first time, to measure the 3D relativistic plasma dynamics deep within a solid, primarily driven by energetic electrons released during relativistic laser-plasma interactions, spanning from a few 100 femtoseconds to over tens of picoseconds. When using plasma as an optical tool, the ability of this characterization method not only allows the mitigation of the spatio-temporal distortion in femtosecond pulses but also offers the ability to control the dynamics of the plasma. Various highly sensitive applications probed from the front/rear side of the target, such as relativistic high-harmonic generation or ion acceleration, can benefit from this technique by gaining invaluable, hitherto unobserved information about spatial and temporal plasma properties. The versatility of this technique extends beyond its current application, as it also captures the 3D spatio-temporal characteristics of pulses from high-power laser systems spanning the terawatt to petawatt range, all in a single shot, potentially boosting high-intensity laser pulse technology.\\

\section{Methods}
\subsection{Reconstruction procedure of spectral and temporal phase}
\textbf{Measuring spectral phase $\varphi (\omega,\textbf{r})$}: The spectral phase is retrieved from the measured spectral interference using an FFT-based algorithm \cite{Takeda1982}, which follows three simple steps and is depicted in Fig. \ref{fig: Fig1}(b-g). First, a fast Fourier transform is applied to the fringe data (Fig. \ref{fig: Fig1}c), revealing two AC side peaks and a DC central peak. Subsequently, one of the side peaks is filtered out, and an inverse Fourier transform (IFFT) is performed, as shown in the inset in Fig. \ref{fig: Fig1}d. The imaginary part of the logarithm of IFFT represents the spectral phase. The local group delay ($\tau$) is the slope of the measured spectral phase. The second and higher-order terms in Eq. \ref{eq:1} are retrieved by fitting a polynomial to the difference between the measured $\varphi (\omega)$ and the first-order term. The carrier-envelope phase $\varphi (\omega_0)$ is obtained by fitting the numerically reconstructed fringes (with all other calculated terms in Eq. \ref{eq:1}) to the raw spectral fringes, as shown in Fig. \ref{fig: Fig1}e. The accuracy of the numerical code was tested through its application to an extensive dataset of synthetic numerical data, which included different kinds of STC coupling, higher-order distortion in the temporal profile of the pulse, as well as low signal-to-noise.\\

\textbf{Measuring temporal phase $\phi$(t)}: The temporal profile of the probe pulse is given by 
\begin{equation}  \label{eq:4}
    E_{pr}(t) = E_{ref}(t)e^{i\Delta\phi(t)}
\end{equation}
where $E_{ref}$ is the field of the reference pulse which is unperturbed by the pump-induced plasma, and where $\Delta \phi$(t) is the modified temporal phase of the probe beam. It assumes that the absorption of the plasma during the probe pulse is constant. Eq. \ref{eq:4} can be solved to yield the time-domain phase shift as
\begin{equation}\label{eq:5}
    \Delta\phi(t) = -i\text{ln}\left(\frac{F(|E_{pr}(\omega)|e^{i\varphi(\omega)+i\varphi_r(\omega)-i\omega\tau})}{F(|E_{ref}(\omega)|e^{i\varphi_r(\omega)-i\omega\tau})} \right)
\end{equation}

where F{} denotes the Fourier transform with respect to frequency, |$E_{pr}$($\omega$)| and |$E_{ref}$($\omega$)| are the square root of the perturbed and unperturbed probe spectrum, respectively, $\varphi(\omega)$ is the spectral phase shift of the probe induced by the pump, and $\varphi_r(\omega)$ is the spectral phase of the unperturbed probe. Of these quantities, |$E_{pr}$($\omega$)| and $\varphi(\omega)$ are retrieved from the measured spectral fringes, while |$E_{ref}$($\omega$)| and $\varphi_r(\omega)$ is obtained from an independent FROG measurement (Fig. S4 in Supplementary Information).

\subsection{Experimental setup}
The experiment was performed using the 150 TW, 25 fs, 800 nm laser system at the Tata Institute of Fundamental Research (TIFR), Mumbai. Fig. S3 of the Supplementary Information shows both the measured picosecond temporal contrast and the femtosecond temporal profile of the pump pulse. These p-polarized laser pulses using a reduced 100 mJ energy were focused on the target with an f/3 off-axis parabolic (OAP) mirror at 40° incidence, corresponding to a laser spot size of 10 µm and a focused peak intensity of 3× $10^{18}$ W/cm$^2$ (Fig. 3a).

A small fraction (5$\%$) of the main laser beam was extracted and up-converted to its second harmonic (SHG) by a 2 mm-thick $\beta$-barium borate (BBO) crystal. This up-converted beam was then employed as a probe pulse. The measured temporal profile of the probe pulse is shown in Fig. S4 of the Supplementary Information. The probe pulse is focused on the rear side of the target with a spot size of 80 $\mu$m. Both the pump and probe spots were measured using a magnified long-working-distance objective lens coupled to a CCD camera. X-ray measurements were conducted using a sodium iodide (NaI) detector to ensure that the target was in the focal plane of the pump beam. A high-precision motorized retro-reflector delay stage (from PI Physik Instrumente) was utilized to introduce a time delay in the probe pulse relative to the pump pulse. Temporal alignment between the pump and probe pulses was achieved by observing a sharp increase in the reflectivity of the probe pulse from the plasma generated by the pump pulse as the probe pulse was scanned in time relative to the pump pulse, as depicted in Fig. S5 in the Supplementary Information.

The reflected probe pulse from the target was collected by an f/2 plano-convex lens and imaged on the lenslet array of our STC measuring device using a set of two plano-convex lenses (Fig. 3a). To correctly locate the imaging plane, we mounted a resolution test target (1951 USAF Target from Thorlabs) in the target plane, which was then imaged in the plane of the lenslet array with a resolution of $\sim$ 6 microns. A portion of the reflected probe pulse is also sent to a photodiode to monitor the shot-to-shot spatial matching between the pump and probe pulses. To block the pump and plasma noise, BG-39 and interference filters centered at 400 nm were used. The experiment was carried out in a vacuum chamber at $10^{-5}$ Torr.\\

\bibliography{Main}
\bibliographystyle{ieeetr}

\vspace{10mm}

\noindent\begin{Large}
\textbf{Acknowledgements}\end{Large}
GRK acknowledges partial support from J.C. Bose Fellowship grant (JBR/2020/000039) from the Science and Engineering Board (SERB), Government of India. ADL acknowledges the partial support from the Infosys-TIFR Leading Edge Research Grant (Cycle 2). LV acknowledge the support from Vetenskapsr\aa det (2019-02376) and (2020-05111). We acknowledge Sunil B. Shetye of the TIFR Technical Services for his invaluable help in setting up the experiment.\\

\noindent\begin{Large}
\textbf{Author contributions}\end{Large}
 GRK conceived the idea and supervised the whole study. A.D. designed the setup, performed the experiment with help from A.D.L., C.A., A.C., and Y.M.V. A.D. analysed the data. The results and interpretation were discussed and finalized by A.D., L.V., and G.R.K.  A.D. wrote the first draft of the manuscript and finalized the manuscript together with L.V. and G.R.K. All authors contributed to the manuscript.\\

\noindent\begin{Large}
\textbf{Competing interests}\end{Large}
The authors declare no competing interests.\\

  \noindent\begin{Large}
\textbf{Data availability}\end{Large}
The data that support the findings of this study are available from the corresponding author upon reasonable request.\\

\noindent\begin{Large}
\textbf{Additional information}\end{Large}
Supplementary information is available for this paper. Correspondence and requests for materials should be addressed to A.D. and G.R.K.\\

\end{document}



\title{Single-Shot, Spatio-Temporal Metrology of Relativistic Plasma Optics - Supplementary Information} 



\author{Ankit Dulat}
\affiliation{Tata Institute of Fundamental Research, 1 Homi Bhabha Road, Colaba, Mumbai 400005, India}

\author{Amit D. Lad}
\affiliation{Tata Institute of Fundamental Research, 1 Homi Bhabha Road, Colaba, Mumbai 400005, India}

\author{C. Aparajit}
\affiliation{Tata Institute of Fundamental Research, 1 Homi Bhabha Road, Colaba, Mumbai 400005, India}

\author{Anandam Chaudhary}
\affiliation{Tata Institute of Fundamental Research, 1 Homi Bhabha Road, Colaba, Mumbai 400005, India}

\author{Yash M. Ved}
\affiliation{Tata Institute of Fundamental Research, 1 Homi Bhabha Road, Colaba, Mumbai 400005, India}

\author{Laszlo Veisz}
\affiliation{Department of Physics, Umeå University, SE-90187 Umeå, Sweden}

\author{G. Ravindra Kumar}
\affiliation{Tata Institute of Fundamental Research, 1 Homi Bhabha Road, Colaba, Mumbai 400005, India}


\pacs{}

\maketitle 

\makeatletter
\renewcommand \thesection{S\@arabic\c@section}
\renewcommand\thetable{S\@arabic\c@table}
\renewcommand \thefigure{S\@arabic\c@figure}
\makeatother

\textbf{Content}

\vspace{5 mm}

\textbf{S1.} Experimental Setup for Measuring STC Effects Induced by a Fresnel Biprism

\vspace{5 mm}

\textbf{S2.} Characterization of the pump and probe pulses used in the experiment

\vspace{5 mm}

\textbf{S3.} Modeling the Effect of Plasma-Induced Spatio-temporal Distortion on the Spectral Profile of the Pulse

\vspace{5 mm}

\textbf{S4.} Temporal matching of pump and probe pulses

\vspace{5 mm}

\textbf{S5.} Effect of temporal phase modulation on spectral profile of probe pulse

\vspace{5 mm}

\textbf{S6.} Pump-probe sechamatic diagram

\vspace{5 mm}

\textbf{S7.} References

\newpage

\section{\NoCaseChange{Experimental Setup for Measuring STC Effects Induced by a Fresnel Biprism }} \label{supp: Fresnel-Biprism}

\begin{figure*}[h]
\centering
\includegraphics[width=0.7\linewidth]{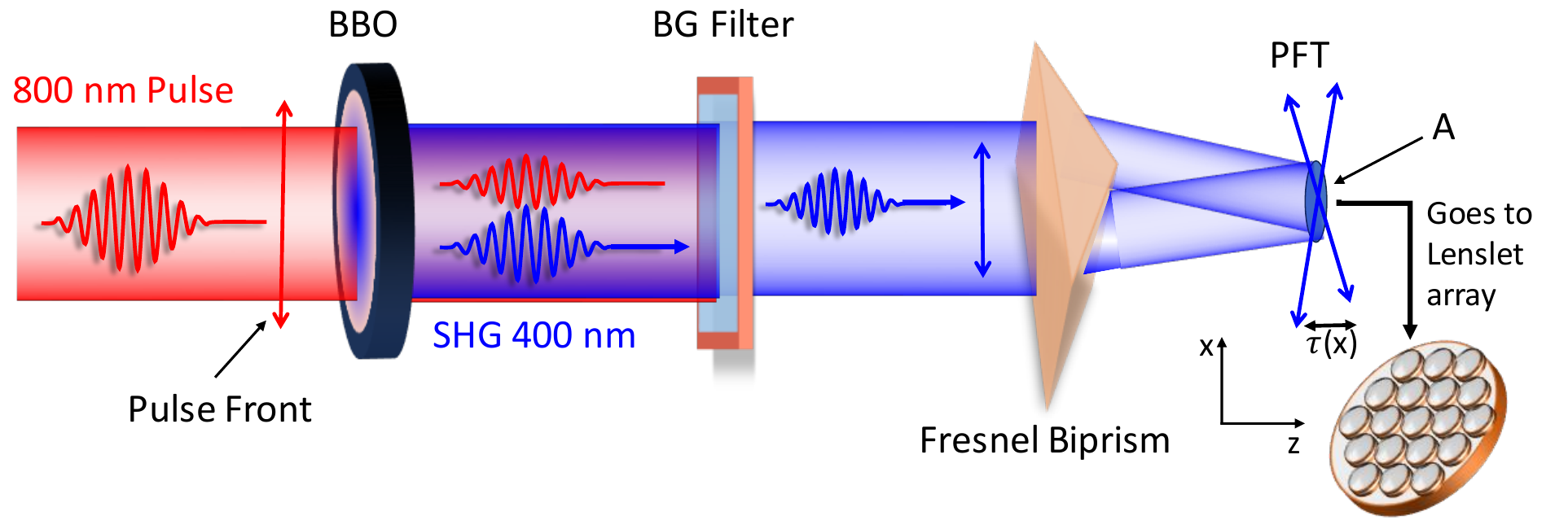}
\caption{\textbf{Experimental setup used for characterizing the pulse front distortion induced by the Fresnel biprism} An 800 nm femtosecond pulse from a Ti:Sapphire laser was upconverted to its second harmonic (400 nm, 260 fs pulse width). To remove the unconverted 800 nm light, a Blue-Green (BG-39) filter was inserted into the optical path. The resulting beam was incident normally on a Fresnel biprism (with an apex angle of 170 degrees). The output pulses from the two spatial halves of the biprism were measured at the intersection plane labeled 'A', using the technique defined in the main manuscript.}\label{fig: FigS1}
\end{figure*}

\begin{figure*}[h]
\centering
\includegraphics[width=0.7\linewidth]{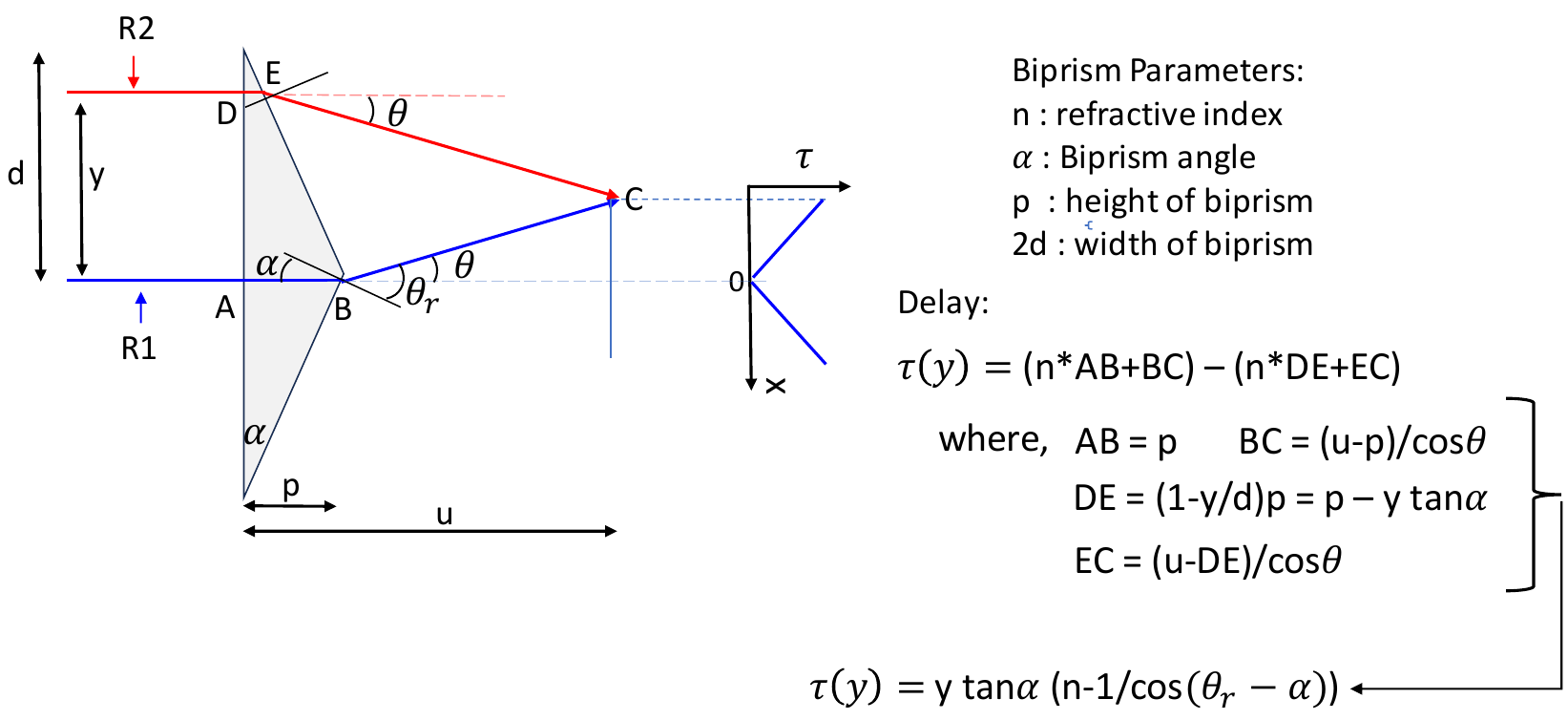}
\caption{Schematic diagram for calculating the maximum delay induced by the Fresnel Biprsim due to pulse front tilt. The blue and red rays correspond to the maximum path difference created by the biprism.}\label{fig: FigS2}
\end{figure*}

The relative group delay introduced by the Fresnel biprism between two halves of the beam can be calculated using the schematic diagram given in Fig. \ref{fig: FigS2}. 

\section{\NoCaseChange{Characterization of the pump and probe pulses used in the experiment}} \label{supp: pulse-characterization}

\begin{figure*}[!h]
\centering
\includegraphics[width=0.7\linewidth]{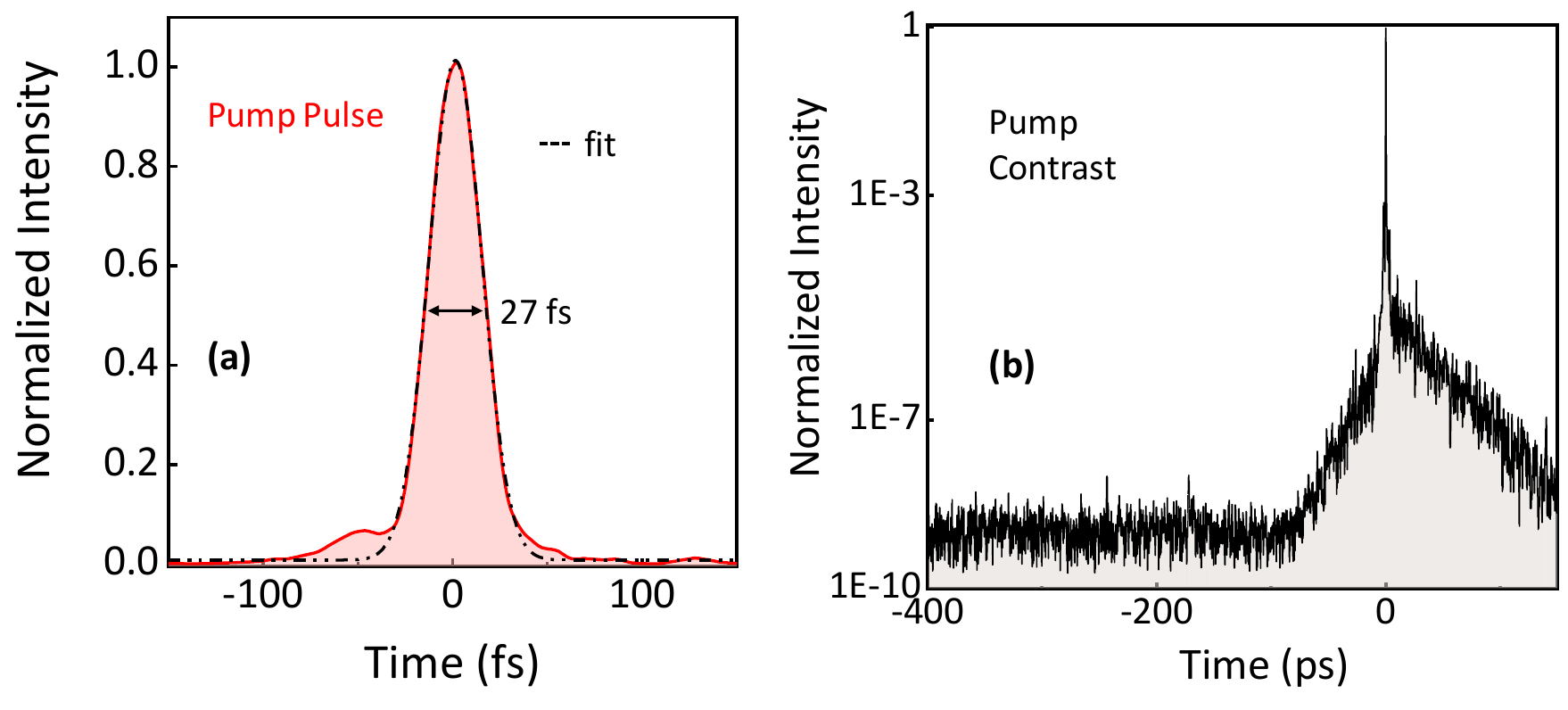}
\caption{\textbf{The temporal profile of pump pulse characterized using SPIDER \cite{Iaconis:98} and SEQUOIA \cite{ALBRECHT198159,Eckardt}.} \textbf{a} Shows the femtosecond temporal profile of the pulse. \textbf{b} The picosecond contrast of the pulse. }\label{fig: FigS3}
\end{figure*}

\begin{figure*}[ht]
\centering
\includegraphics[width=0.7\linewidth]{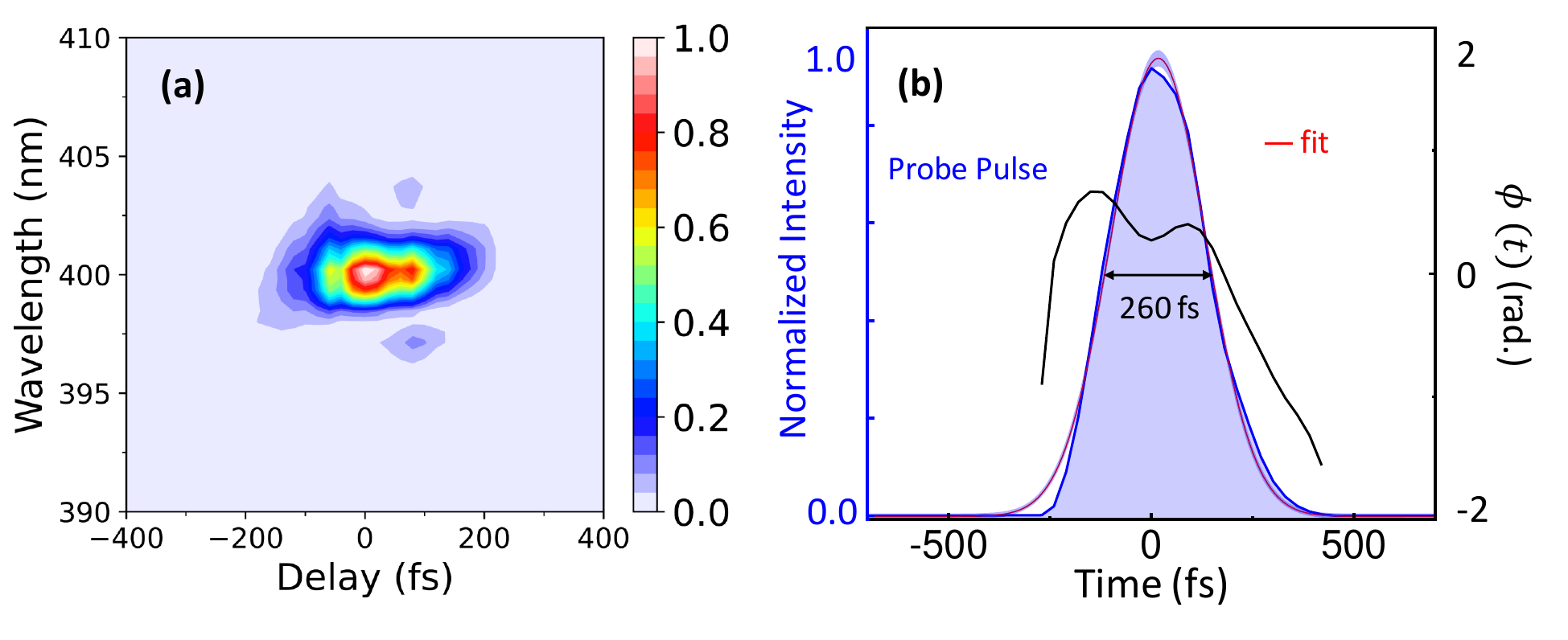}
\caption{\textbf{The temporal profile of the second harmonic probe pulse measured using SD-FROG \cite{Trebino-Kane}.} \textbf{a} Shows the measured FROG trace of the probe pulse. \textbf{b} Reterived femtosecond temporal profile of the pulse}\label{fig: FigS4}
\end{figure*}

\newpage

\section{\NoCaseChange{Modeling the Effect of Plasma-Induced Spatio-temporal Distortion on the Spectral Profile of the Pulse}} \label{supp: Reconstruction procedure}

\subsection{The effect of plasma mirror acceleration on the bandwidth}

The electric field of the reflected probe from the plasma can be calculated within the Fresnel approximation \cite{Dulat:22} because the high-contrast pump pulses presented here have a pre-plasma scale length of L/$\lambda \ll 1$ . However, this approximation breaks down as the plasma starts expanding, and within a few picoseconds (ps), the scale length could extend to a few microns. Initially, we consider the case where L/$\lambda \ll 1$, which is suitable for describing early-time evolution. Later in this section, we will discuss how the results are affected when this approximation no longer applies. For further simplification, we assume that the reflectivity of the plasma doesn’t change significantly during its interaction with the probe pulse. 
In the experiment, the probe pulse passes through approximately 3 cm of glass, resulting in a positive linear chirp denoted as $a$. Based on this simplified plasma mirror model, the electric field of the reflected light can be expressed as:

\begin{equation} \label{eq:1}
E_r(t) = r_0E_0e^{-(\frac{t-t_0}{\tau_L})^2}e^{j(\omega_0(t-t_0)+kx(t)+a(t-t_0)^2)}
\end{equation}
where $r_0$ is the reflectivity of the plasma, $\tau_L$ is the pulsewidth of the incident probe, and $x(t)$ represents the position of the plasma surface at time t. Utilizing this simple model, we can express $x(t) = vt + \frac{1}{2}bt^2$, where $v$ and $ b$ represent the instantaneous velocity and acceleration of the plasma surface, respectively. The spectral profile of the reflected probe pulse can be obtained via the Fourier transform of the reflected field as given by Eq. \ref{eq:1}

\begin{equation} \label{eq:2}
\mathcal{E}_r(\omega) = F[E_r(t)] \simeq r_0E_0\int^{\infty}_{-\infty}e^{-(\frac{t-t_0}{\tau_L})^2}e^{j((\omega_0-\omega)(t-t_0)+a(t-t_0)^2+\frac{\omega_0}{c}vt+\frac{1}{2}b\frac{\omega_0}{c}t^2)}dt
\end{equation}

This integral can be simplified to yield spectrum \cite{Sauerbrey1996}:
\begin{equation} \label{eq:3}
    S(\omega) \sim \text{exp}[-(\omega-\omega^1)^2/(2\Delta^2)] \text{, \hspace{3 mm} where central frequency is } \omega^1 = \omega_0\left(1+\frac{v}{c}+\frac{bt_0}{c}\right) 
\end{equation}

\begin{equation} \label{eq:4}
   \text{and bandwidht is given by \hspace{3mm}} \Delta^2 = 1+(\alpha+\beta)^2 \text{\hspace{3 mm} where  } \alpha = a\tau_L^2 \text{ ,\hspace{3 mm}}\beta = \omega_0b\tau_L^2/(2c)
\end{equation}

The parameters $\alpha$ and $\beta$ denote the initial chirp and the instantaneous acceleration of the plasma surface, respectively, both in normalized units. As indicated in Eq. \ref{eq:4}, when an unchirped probe pulse ($\alpha$ = 0) reflects from an accelerating plasma ($\beta \neq$ 0), the bandwidth will always increase. On the other hand, when the probe pulse has a positive chirped ($\alpha$ > 0) and the plasma accelerates by the radiation pressure of the pump pulse inward, i.e., ($\beta$ < 0), the spectral width can decrease because its effective chirp decreases. However, when both $\alpha$ and $\beta$ > 0, the bandwidth of the probe pulse can be increased further.

However, when the scale length of the plasma is much larger than the $\lambda$, then the bandwidth of the probe pulse can also be affected by the phase modulation induced in the probe due to plasma nonlinearities; one such example is self-phase modulation. So one needs to be careful while calculating the contribution of each effect in those cases.

\section{\NoCaseChange{Temporal matching of pump and probe pulses}}
At negative time delay, the probe pulse is ahead of the pump and hence experiences low reflectivity of the glass target. After t=0, there is a rapid increase in reflectivity (rise time < 100 fs). This is due to the rapid ionization of the fused silica target by the pump pulse, hence it acts as a plasma mirror for the probe pulse. After reaching the peak, the reflectivity begins to drop slowly because of the absorption of the probe pulse in the expanding plasma.

\begin{figure*}[h]
\centering
\includegraphics[width=0.4\linewidth]{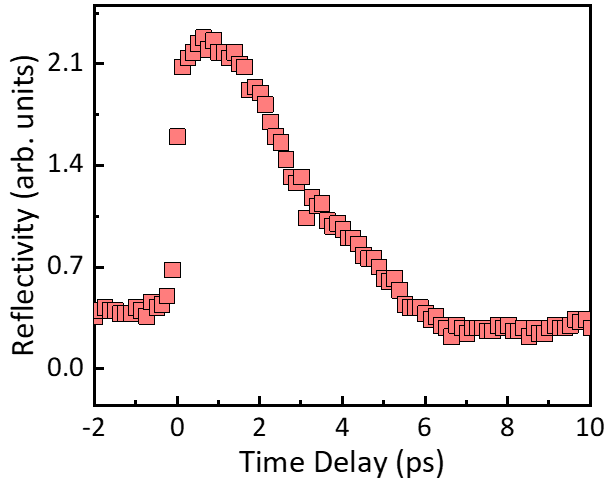}
\caption{Time-resolved reflectivity of the probe pulses from  a fused silica target that is excited by a pump pulse of intensity $\sim 10^{15}$~W/cm$^2$}\label{fig: FigS5}
\end{figure*}

\section{\NoCaseChange{Effect of temporal phase modulation on the spectral profile of probe pulse}}
In this section, we numerically check the effect of temporal phase modulation induced by the hydrodynamically expanding plasma on the spectral profile of the probe pulse. We also check the contribution of higher-order terms in the Taylor series expression to the temporal phase, given by Eq. \ref{eq:5}.
\begin{figure*}[h]
\centering
\includegraphics[width=1\linewidth]{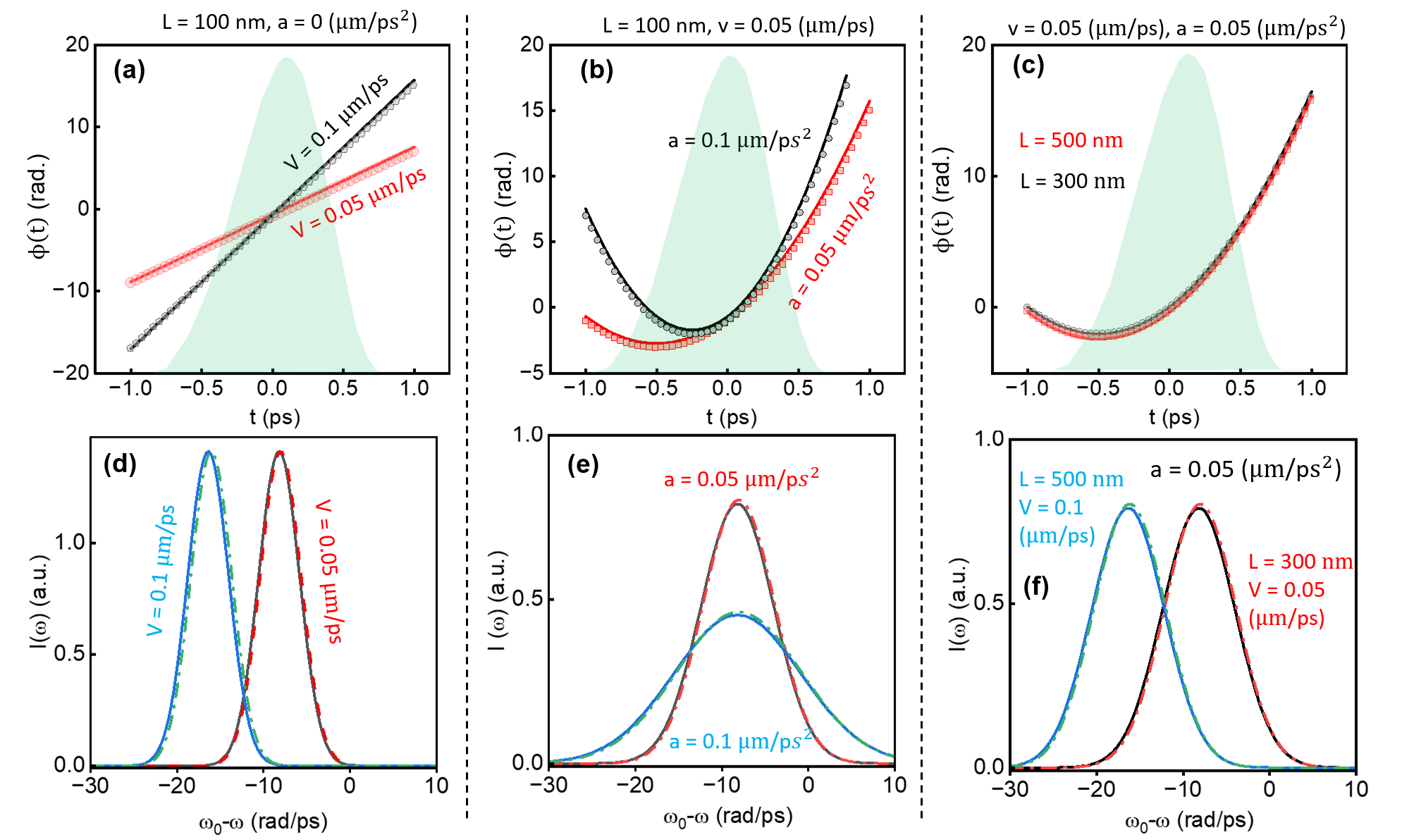}
\caption{(\textbf{a-c}) depicts the modulation in the temporal phase experienced by the probe pulse as it interacts with expanding plasma with an initial scale length of 100 nm (in \textbf{a,b}); \textbf{a} Compare $\phi(t)$ for two different velocities of the critical surface and \textbf{b} compare it for two different acceleration values. \textbf{c} Compares the phase experienced for two different scale lengths of the plasma. The curves with scatter plots are obtained using only the first two terms in the Taylor series expansion of Eq. \ref{eq:5}, while the solid curves are without any approximation. \textbf{d-f} shows the corresponding change in the spectral profile of the probe pulse. \textbf{d} Doppler shift due to linear phase modulation. \textbf{e} Bandwidth broadening and Doppler shift due to the accelerating critical surface of the plasma. \textbf{f} Comparison of the Doppler shift and the change in the bandwidth of the probe pulse for two different scale lengths of the plasma. The curves with dashed line plots are obtained using only the first two terms in the Taylor series expansion of Eq. \ref{eq:5}, while the solid curves are without any approximation.}\label{fig: FigS6}
\end{figure*}

\begin{equation} \label{eq:5}
\phi(t,\textbf{r})  = 2\int^{z=z_{c}(t)}_{z_R} \frac{\omega}{c}\sqrt{\epsilon(\textbf{r},s)} ds  \approx \frac{2\omega_0}{c}\left (\int^{z=z_{c}(t)}_{z_R}ds - \frac{1}{2n_{c}}\int^{z=z_{c}(t)}_{z_R}n_{e}(\textbf{r},s,t)ds\right)+  \psi[f(n_e/n_c)]
\end{equation}

where $z_R$ is a reference point far from the target, $z_c(t)$ is the position of the reflecting surface at time t, \textbf{r} is the coordinate along the transverse spatial profile and s is along the beam path. $\epsilon$ is the permittivity that can be approximated as $\epsilon$=1-(n$_{e}$/n$_{c}$), with n$_{e}$ and n$_{c}$ being the local electron density and critical density, respectively. $\psi$ is the function consisting of the integral of higher-order terms in the Taylor series expansion of $\epsilon$. Assuming exponential plasma density profile $n_e = n_0 exp(-z/L)$, where L is the scale length of the plasma.\\
The critical surface of the plasma is expanding with velocity \textbf{v} and acceleration \textbf{a}. A 400 nm probe pulse of 260 fs pulse duration interacts with the plasma. The modulation in the temporal phase experienced by the probe pulse is depicted in Fig. S6 (a–c), and the corresponding change in the spectral profile of the pulse is shown in Fig. S6 (d–f). The plasma parameters relevant to this study are used in the calculations, and shown as the title and in the inset of each figure. In each of these figures, the curves with scatter plots or dashed lines are obtained using only the first two terms in the Taylor series expansion of Eq. \ref{eq:5}, while the curves with solid lines are without approximation. We observe that higher-order terms in Eq. \ref{eq:5} do not contribute significantly for an expanding solid-density plasma and can be neglected for the plasma parameter discussed in this paper.

\newpage
\section{\NoCaseChange{Pump-probe sechamatic diagram}}

\begin{figure*}[h]
\centering
\includegraphics[width=0.5\linewidth]{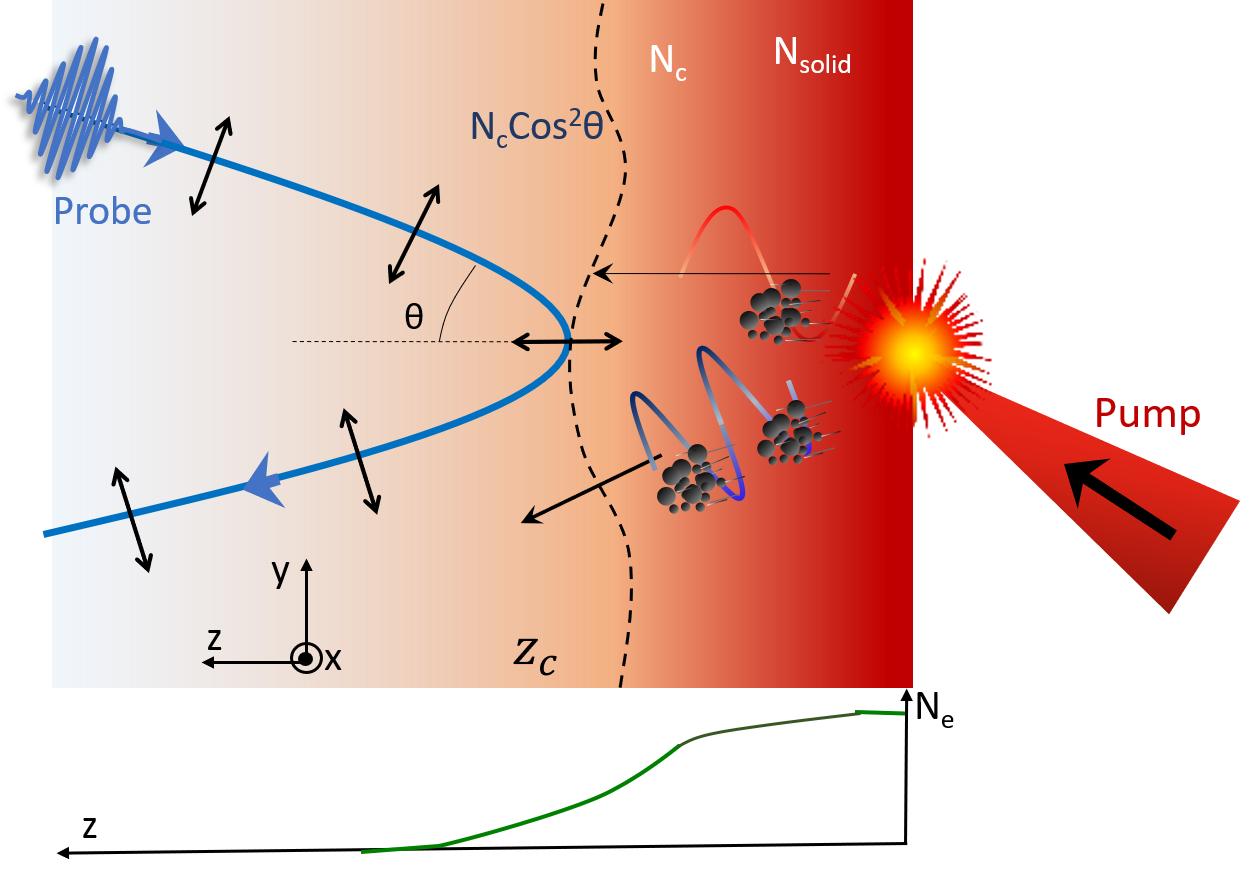}
\caption{Intense pump pulse interaction with the front surface of a metal-coated glass target produces energetic electrons that go deep inside the target and deposit their energy via ionizing the solid, which is being probed by the probe pulse coming from the rear side.}\label{fig: FigS7}
\end{figure*}

\section{\NoCaseChange{References}}
\bibliography{Supplementary}